\documentclass[modern]{aastex631}

\usepackage{hyperref}
\let\tablenum\relax % Issue with siunitx/AASTeX
\usepackage{siunitx}
\usepackage{graphicx}%
\usepackage{multirow}%
\usepackage{amsmath,amssymb,amsfonts}%
\usepackage{amsthm}%
\usepackage{mathrsfs}%
\usepackage[title]{appendix}%
\usepackage{xcolor}%
\usepackage{textcomp}%
\usepackage{booktabs}%
\usepackage{algorithmicx}%
\usepackage{algpseudocode}%
\usepackage{listings}%
\usepackage{upgreek}

\begin{document}

\title{MultIHeaTS: a Fast and Stable Thermal Solver for Multilayered Planetary Surfaces}

\correspondingauthor{Cyril Mergny}

\author[0009-0002-1910-6991]{Cyril Mergny}
\affiliation{Universit\'e Paris-Saclay, CNRS, GEOPS \\
Orsay, 91405, France}
\email{cyril.mergny@universite-paris-saclay.fr}

\author[0000-0002-2857-6621]{Fr\'ed\'eric Schmidt}
\affiliation{Universit\'e Paris-Saclay, CNRS, GEOPS \\
Orsay, 91405, France}
\affiliation{Institut Universitaire de France \\
Paris, France}

\begin{abstract}
A fully implicit scheme is proposed for solving the heat equation in 1D heterogeneous media, available as a computationally efficient open-source Python code. The algorithm uses finite differences on an irregular grid and is unconditionally stable due to the implicit formulation. The thermal solver is validated against a stiff analytical solution, demonstrating its robustness in handling stiff initial conditions. Its general applicability for heterogeneous cases is demonstrated through its use in a planetary surface scenario with non-linear boundary conditions induced by black body thermal emission. MultIHeaTS advantageous stability allows for computation times up to 100 times faster than Spencer's explicit solver, making it ideal for simulating processes on large timescales.
This solver is used to compare the thermal signatures of homogeneous and bilayer profiles on Europa.  Results show that homogeneous materials cannot reproduce the thermal signature observed in bilayer profiles, emphasizing the need for multilayer solvers. 
In order to optimize the scientific return of space mission, we propose a strategy made of three local time observations that is enough to identify a bilayer media, for instance for the next missions to the Jovian system.
A second application of the solver is the estimation of the temperature profile  of Europa's near surface (first 10 meters) throughout a one-million-year simulation with varying orbital parameters. The probability distribution of temperature through depth is obtained. 
Among its various applications, MultIHeaTS serves as the core thermal solver in a multiphysics simulation model detailed in the companion article  \citet{Mergny2024i}.

%\textit{LunaIcy: Exploring Europa's Icy Surface Microstructure through Multiphysics Simulations}.
%By comparing measured brightness temperature patterns with such solver outputs, valuable surface property information can be retrieved, providing valuable insights for remote sensing missions.

\end{abstract}


\section{Introduction}

The heat equation is a fundamental partial differential equation that governs the behavior of heat transfer in various physical systems. However, when dealing with heterogeneous media, where the thermal properties of the material vary spatially, this equation becomes too complex to solve analytically \citep{Jaeger1950, Loeb2019}. In one dimension, the heat equation for conduction transfer can be expressed as:
\begin{equation}
    \rho(x, t) c_{\mathrm{p}}(x, t) \dfrac{\partial T(x, t)}{\partial t} =  \dfrac{\partial}{\partial x} \left( k(x, t) \dfrac{\partial}{\partial x} T(x, t) \right)
    +  Q(x, t)
    \label{eq:heat-equation}
\end{equation}
where $x$ denotes space and $t$ denotes time, $T$ is the temperature, $\rho$ is the density, $c_{\mathrm{p}}$ is heat capacity, $k$ is thermal conductivity, and $Q$ denotes an additional and optional source or sink term.
We propose to solve numerically Equation \ref{eq:heat-equation} where all thermal properties can vary continuously over space and time, and with non-constant time and space increments. 
For the near surface, the vertical temperature gradients induced by the input solar flux is often much larger than any lateral heat variations, justifying the use of a 1D model for many planetary science applications. At greater depths (hundreds of meters), a 3D model would be beneficial due to reduced solar flux influence, but given the lack of knowledge on lateral heat transfer of icy moons, the 1D model will suffice.

Numerous numerical methods have been developed in various fields to solve the 1D heat equation for heterogeneous media. These methods include finite element and finite difference approaches \citep{Roubicek1990, Lage1996, Nissen2017, Loeb2019, Masson2020}. However, in planetary sciences, widely used solvers such as Thermprojrs \citep{Spencer1989}, MARSTHERM \citep{Putzig2007}, KRC \citep{Kieffer2013}, and Heat1D \citep{Hayne2015}, rely on explicit finite difference schemes which are known to have instability issues \citep{Press1992}. While there are semi-implicit solvers, most notably Schorghofer's Planetary Code Collection \citep{Schorghofer2022} \footnote{Schorghofer's Planetary Code Collection was developed between 2001 and 2003.} and later \cite{Young2017}, none have pursued a fully implicit derivation, despite its known numerical stability. Our study stands out as the first to take the approach of a fully implicit scheme in the context of planetary thermal modeling.

The target science case is in Earth and Planetary science with surface conditions subject to solar illumination. In this scenario, the upper boundary condition (i.e., the flux leaving the surface) is determined by the energy equilibrium between the input solar flux and the gray body emission from the surface \citep{Spencer1989}.
\begin{equation}
    \forall t, k(0,t) \left. \dfrac{\partial T(x, t)}{\partial x}\right|_{x=0} =  - F_{\mathrm{solar}}(t) + \epsilon \cdot \sigma_{\mathrm{SB}} \cdot T(0, t)^4
    \label{eq:energy-eq}
\end{equation}
where $\epsilon$ is the thermal emissivity and $\sigma_{\mathrm{SB}}$ the Stefan-Boltzmann constant.
A simplified model for solar illumination at the equator with zero obliquity throughout a sidereal day can be expressed as a truncated sinusoidal function \citep{Spencer1989} :
\begin{equation}
	F_{\mathrm{solar}}(t) = \left\{ 
		\begin{array}{ll}
			\left(1 - A \right) \dfrac{G_{\mathrm{SC}}}{d^2} \cos \left(\dfrac{2 \pi t}{P} \right) & \text{ if } {2 \pi t}/{P} \text{ in } \left[-\frac{\pi}{2}, \frac{\pi}{2}\right] \\ 
		\text{or} \\
		0   &  \text{ if }  {2 \pi t}/{P} \text{ in } \left]\frac{\pi}{2}, \frac{3\pi}{2}\right[ \\
		\end{array}
	\right.
 \label{eq:slr_flux_model}
\end{equation}
where $A$ is the surface albedo, $G_{\mathrm{SC}}$ the solar constant in W.m$^{-2}$ and $d$ the distance between the Sun and the planet in AU. 
While this simplified model captures the essential features of solar illumination, more realistic estimations, for instance from NASA's SPICE toolkit \citep{Acton1996, Acton2018} can be easily integrated to account for different inclinations, eccentricities and more, as shown in Section \ref{sec:application}.

In the planetary science context, various numerical models have been developed to study the behavior of both homogeneous surfaces \citep{Wesselink1948, Rozitis2011, Kieffer2013}, and heterogeneous media, as shown in Spencer's implementation of the explicit Euler scheme \citep{Spencer1989}. 
Although the explicit method is easier to numerically implement and could have a faster computational time per iteration, its main drawback is its conditional stability \citep{Press1992}.
A Crank-Nicolson scheme \citep{Crank1947} has been implemented for planetary science cases by \citet{Schorghofer2022}. This scheme is more complex to implement and requires the inversion of tridiagonal matrix. In return, it offers the advantage of being unconditionally stable and more precise than first-order explicit and fully implicit schemes for homogeneous media, owing to its second-order truncation error in time \citep{Mazumder2016}.  However, the development of a fully implicit algorithm is also advantageous, as it can easily account for spatial  heterogeneities by staying stable \citep{Press1992, Oesterby2003, Langtangen2017}.

 Although the implicit Euler scheme is also unconditionally stable, to our knowledge, there are no codes available online that use it to solve the heat equation in a heterogeneous medium for planetary surfaces. 
 Therefore, our goal is to provide the scientific community with an open-source, easy-to-use, and versatile fully implicit solver called MultIHeaTS (Multi-layered Implicit Heat Transfer Solver) for solving the heat equation in such conditions. The source code is available online\footnote{MultIHeaTS open-source code is available at the IPSL Data Catalog:\dataset[10.14768/9763d466-db02-4f29-8ad5-16e6e0187bd4]{\doi{10.14768/9763d466-db02-4f29-8ad5-16e6e0187bd4}}}. Even though our target science is planetary science with surface conditions subject to solar illumination \citep{Wesselink1948, Spencer1989, Schorghofer2022, Rozitis2011, Kieffer2013}, this approach is applicable in a wide range of boundary conditions for a large set of scientific and technical cases.

When thermal properties depend on temperature, they introduce non-linearity into the heat equation, making it challenging to solve, regardless of the numerical scheme. For instance, both thermal conductivity $k(T)$ and heat capacity $c_p(T)$ exhibit strong temperature dependence when considering conditions on silicate bodies within a distance of less than $\SI{3}{AU}$ from the Sun \citep{Watson1964, WoodsRobinson2019}.
Nevertheless, by linearizing the temperature-dependent thermal properties, we can effectively solve the modified heat equation. This approach, used for example in Heat1D \citep{Hayne2015} and \cite{Schorghofer2022} expresses the thermal properties as functions of the previous temperature, allowing iterative updates at each time step using time-dependent variables $k(t), c_p(t)$, and $\rho(t)$.
It is important to note that the linearization process may impact numerical stability, particularly in situations involving significant temperature variations. While further stability analysis would be advised for future studies, using smaller time steps and finer linearization (like discussed in Section \ref{sub:nonlinearBC}) will solve the stability issues.

In the first section of this article, we begin by presenting the mathematical derivation of the heat equation in heterogeneous media using the implicit Euler method. We then proceed to validate the MultIHeaTS algorithm by comparing its results to analytical solutions and other existing algorithms. Then, our thermal model is used to investigate the thermal signatures of bilayer profiles on Europa, providing valuable insights that can guide the timing of measurements for upcoming missions. Finally, the thermal model is applied to a one million year simulation of Europa's temperature profile, using a precise orbital description.

\section{Methods}

To discretize the heat equation, we used finite differences on an irregular spatial grid consisting of $n_x$ points, which we iterated for a total of $n_t$ iterations. The spatial and temporal parameters were discretized as follows:
\begin{equation}
\begin{cases}
    x \rightarrow x_n &= x_{n-1} + \Delta x_n \\
    t \rightarrow t^i &= t^{i-1} + \Delta t^i .\\
    
\end{cases}
\end{equation}
Here, $n$ is an integer such that $n \in \{0, \: \dotsc \: , n_x-1 \} $, representing the $n$th element in the spatial dimension, and $i$ is an integer such that $i \in \{0, \: \dotsc \: , n_t-1 \} $, representing the $i$th element in the time dimension. It is worth noting that both the spatial $\Delta x_n$ and temporal $\Delta t^i$ increments may not be constant respectively over the spatial and temporal grid.

\subsection{Backward Euler Finite Differences on an Irregular Grid}

Following the method described in \citep{Sundqvist1970}, the first order derivative of function $f$ using the central difference approximation can be written as :
\begin{equation}
\frac{\partial f}{\partial x} (x_n) = \frac{f_{n+1}- f_{n}}{2 \Delta x_n} + \frac{f_{n}- f_{n-1}}{2 \Delta x_{n-1}} + \mathcal{O}(\Delta x_n^2)
 \label{eq:1_deriv}
\end{equation}
and the second order derivative as :
\begin{equation}
	\frac{\partial f^2}{\partial x^2} = \dfrac{2 \left( f_{n+1} \Delta x_{n-1} + f_{n-1} \Delta x_{n} - f_n \left( \Delta x_{n} + \Delta x_{n-1}\right) \right) }{\Delta x_{n} \Delta x_{n-1} \left( \Delta x_{n} + \Delta x_{n-1}\right)} + \mathcal{O}(\Delta x_n^3)
 \label{eq:2_deriv}
\end{equation}

By using Equations \ref{eq:1_deriv} and \ref{eq:2_deriv}, the heat equation can be discretized with second order accuracy in space and first order in time. After factorizing each temperature term, the resulting equation is given by:
\begin{align}
\begin{split}
	T_n^i + r_n^i Q_n^i = &T_{n-1}^{i+1}  \left[ \frac{-r_n}{\Delta x_{n-1}} \left( - \frac{1}{2} \frac{\partial k}{ \partial x} + \frac{2 k_n}{\Delta x_n + \Delta x_{n-1}} \right)  \right] \\
	+ &T_{n}^{i+1} \left[ 1 - \frac{r_n}{\Delta x_{n}\Delta x_{n-1}} \left( \frac{\left(\Delta x_{n}- \Delta x_{n-1} \right)}{2} \frac{\partial k}{ \partial x} - 2 k_n \right) \right] \\
      + &T_{n+1}^{i+1}\left[ \frac{-r_n}{\Delta x_{n}} \left( \frac{1}{2} \frac{\partial k}{ \partial x} + \frac{2 k_n}{\Delta x_n + \Delta x_{n-1}} \right) \right]
\end{split}
    \label{eq:dis_Heat}
\end{align}
with $T_n^i$ the temperature of the $n$th cell at time $t^i$, and $r_n$ a coefficient expressed as $r_n^i = \Delta t^i / (\rho_n^i c_{\mathrm{p}, n}^i)$. With matrix notation this is equivalent to the system :
\begin{equation}
    \begin{pmatrix}
    b_0 & c_0 & &  &\hdots  & & 0  \\
    a_1 & b_1 & c_1 &  \\
      &\ddots & \ddots & \ddots     \\
    \vdots  &  & a_n & b_n & c_n & &  \vdots \\
    & & &\ddots & \ddots & \ddots     \\
    & & &  & a_{n_x-2} & b_{n_x-2} & c_{n_x-2} \\
    0 & &  \hdots & & & a_{n_x-1} & b_{n_x-1}  \\
    \end{pmatrix}
    \cdot
    \begin{pmatrix}
    T_{0} \\
    \vdots \\
    T_{n-1} \\
    T_{n}  \\
    T_{n+1} \\
    \vdots \\
    T_{n_x-1}
    \end{pmatrix}^{i+1}
     = 
     \begin{pmatrix}
        s_{0} \\
        \vdots \\
        s_{n-1} \\
        s_{n}  \\
        s_{n+1} \\
        \vdots \\
        s_{n_x-1}
        \end{pmatrix}^{i}
\end{equation}
where the coefficients $a_n$, $b_n$, $c_n$ and $s_n$ are given by
\begin{equation}
\forall n \in \{1, \dots, n_x-2 \} 
\begin{cases}
& a_n =   \dfrac{-r_n}{\Delta x_{n-1}} \left( - \dfrac{1}{2} \dfrac{\partial k}{ \partial x} + \dfrac{2 k_n}{\Delta x_n + \Delta x_{n-1}} \right) \\
& b_n = 1 - \dfrac{r_n}{\Delta x_{n}\Delta x_{n-1}} \left( \dfrac{\left(\Delta x_{n}- \Delta x_{n-1} \right)}{2} \dfrac{\partial k}{ \partial x} - 2 k_n \right) \\
& c_n =  \dfrac{-r_n}{\Delta x_{n}} \left( \frac{1}{2} \dfrac{\partial k}{ \partial x} + \dfrac{2 k_n}{\Delta x_n + \Delta x_{n-1}} \right)\\
& s_n = T_n^i + r_n^i Q_n^i\\
\end{cases}
\label{eq:an_terms}
\end{equation}
and by the boundary conditions for $n=0$ and $n= n_{x}-1$.

\subsection{Linear Boundary Conditions}

MultIHeaTS accepts any type of boundary conditions, including imposed flux or imposed temperatures at the boundaries. A detailed derivation of the Dirichlet boundary conditions can be found in Supplementary Materials Section 1. In the case of planetary surface evolution, we are specifically interested in Neumann boundary conditions, which involve imposed flux. 
In this scenario, the heat flux $\Phi$ is prescribed at the upper boundary ($n=0$) by
\begin{equation}
    \forall t, k_0 \left. \dfrac{\partial T(x, t)}{\partial x}\right|_{x_0} = \Phi_0(t).
    \label{eq:upper_flux}
\end{equation}
By injecting Equation \ref{eq:upper_flux} into the discretized heat Equation \ref{eq:dis_Heat}, the upper boundary condition becomes:
\begin{equation}
    T_0^{i+1} = T_0^i + r_0^i \left[\dfrac{\Phi_0^i }{\Delta x_0 k_0^i} \left(k_{1}^i - 3k_{0}^i\right) +   Q_0^i \right]
      +  2 \dfrac{r_0^i k_0^i}{\Delta x_0^2}  T_{1}^{i+1} - 2 \dfrac{r_0^i k_0^i}{\Delta x_0^2}   T_{0}^{i+1} 
\end{equation}
which gives the expression of the first coefficients of the tri-diagonal matrix 
\begin{equation}
    \begin{cases}
        &b_0 = 1 + 2 r_0 k_0 / \Delta x_0^2  \\
        &c_0 = - 2 r_0 k_0   / \Delta x_0^2 \\
        &s_0 =  T_0 + r_0 \left( \Phi_0 /k_0 \left(k_{1} - 3k_{0} \right)/\Delta x_0 +  Q_0 \right) .\\
    \end{cases}   
    \label{eq:coeff_up}
\end{equation}
The same reasoning can be applied to the bottom boundary condition :
\begin{equation}
    \forall t,  k_{n_x-1} \left. \dfrac{\partial T(x, t)}{\partial x}\right|_{x_{n_x-1}} = \Phi_{n_x-1}(t) 
\end{equation}
%\begin{equation}
%T_N^{i+1} = T_N^i + r_N^i \left[  \Delta x \Phi_N^{i+1} \left(3k_{N}^i - k_{N-1}^i\right)  +  \Delta x^2 Q_n^i\right]
%+  2k_N^i r_N^i T_{N-1}^{i} - 2k_N^i r_N^i T_{N}^{i}
%\end{equation}
which gives the expression of the last tri-diagonal coefficients :
\begin{equation}
    \begin{cases}
        &a_{n_x-1} = - 2 r_{n_x-1} k_{n_x-1} / \Delta x_{n_x-1}^2 \\
        &b_{n_x-1} = 1 +  2 r_{n_x-1} k_{n_x-1} / \Delta x_{n_x-1}^2   \\
        &s_{n_x-1} =T_{n_x-1} + r_{n_x-1} \left[   \Phi_{n_x-1}  / k_{n_x-1} \left( k_{n_x-2} - 3 k_{n_x-1} \right)  /\Delta x_{n_x-1}+  Q_{n_x-1} \right] \\
        
    \end{cases}   
\end{equation}

\subsection{Non-Linear Boundary Conditions}
\label{sub:nonlinearBC}

When dealing with planetary-like surfaces, a non-linearity problem arises from the energy equilibrium described in Equation \ref{eq:energy-eq}. This non-linearity comes from the Stefan-Boltzmann law, which states that the emitted surface flux $\Phi$ is proportional to $T_0^4$, making it a non-linear function of temperature: 
\begin{equation}
    \Phi(t, T_0) \propto T_0^4
\end{equation}

For the explicit scheme, the non-linearity in the surface flux does not pose any issue as the surface flux can be pre-calculated using the previous surface temperature. However, for Crank-Nicolson and Backward Euler schemes, the non-linearity of the Stefan-Boltzmann law makes it impossible to solve the implicit formulation of the upper boundary conditions as given in Equation \ref{eq:coeff_up}. To circumvent this issue, a workaround exist by linearizing the black body emission around a reference temperature $T_{\mathrm{r}}$ \citep{Williams1977, Schorghofer2022}:
\begin{equation}
T_0^{{i+1}^{4}}  = \left( T_{\mathrm{r}} + \delta T \right)^4 \approx T_{\mathrm{r}}^4 + 4 T_{\mathrm{r}}^3 \delta T
\label{eq:linearizeT4}
\end{equation}
where $T_{\mathrm{r}}$ is chosen to be equal to the previous surface temperature $T_0^i$ and  $\delta T = T_0^{i+1} - T_0^{i}$.
%This approach led to the development of a new boundary condition formulation which is sensitive to abrupt changes of energy input, causing the surface temperature to vary rapidly.
%If the reference temperature is far from the actual surface temperature this causes a significant error in the evaluation of the emitted energy. To address this issue, Sch{\"o}rghofer  \citep{Schorghofer2022} uses a predictor-corrector step that iteratively adjusts the reference temperature.

%\textcolor{red}{but it comes with issues \citep{Press1992} ?}

For the MultIHeaTS solver, the boundary surface flux can be calculated either using Equation \ref{eq:linearizeT4} or the previous surface temperature. Our validation tests, discussed in Section \ref{sec:results}, demonstrate that following the same approach as for the explicit scheme does not introduce large errors and the algorithm remains stable.
One advantage of the implicit scheme is that such formulation of the surface flux can be used, while maintaining stability and adding little to no errors \citep{Beam1982}. Unless stated otherwise, this is the method chosen for computing the upper boundary conditions in the rest of this article. We propose that this robustness may be due to the strong damping of oscillations discussed in Section \ref{sec:stability}. 
Similar to the Crank-Nicolson solver, the fully implicit scheme looses its unconditional stability when there are abrupt changes in surface temperatures with such non linear conditions. Nonetheless, the scheme is capable of accurately estimating the upper boundary condition and its domain of stability remains much larger than the explicit solvers as shown in Section \ref{sec:efficiency}.

%iThe two latter are popular since they are unconditionally stable irrespective of
%the magnitude of bμ, and Crank-Nicolson especially so being of second order in time.
%One drawback with Crank-Nicolson, however, is that the solution has a tendency to oscillate
%if there are jump discontinuities in the initial condition or between the initial
%condition and a boundary condition.

%The stability criteria of the explicit scheme is given by the Fourier Number :
%\begin{equation}
%F = \max_{i, n} \left( \dfrac{\alpha_n^i \Delta t^i}{\Delta x^2_n}  \right) < \dfrac{1}{2}
%\end{equation}
%where $\alpha$ is the thermal diffusivity defined as $\alpha = k/\rho c_{\mathrm{p}}$.

\section{Results}
\label{sec:results}
\subsection{Validation with an Analytical Solution}

 Since analytical solutions of the heat equation do not exist in the general case, analytical validation of our model can only be achieved for homogeneous profiles.
%In the case of constant conductivity, density and heat capacity the 1D heat equation can be solved analytically for a given set of initial and boundary conditions.
If we consider a stiff initial condition given by a step function defined on $[0, L]$ by the expression :
\begin{equation}
    \forall x \in [0, L], \, T(x, 0) = \left\{
\begin{array}{ll}
0, & \text{if } x < L/2 \\
1, &  \text{if } x \geq L/2.\\
\end{array}
    \right.
\label{eq:step_fct}
\end{equation}
then the analytical solution with zero-flux boundary conditions can be obtained
through Fourier series decomposition
%rajouter supmat
\begin{equation}
    T_{\mathrm{R}}(x, t) = \frac{1}{2} - \sum_{j=1}^{+\infty} \frac{2}{\pi j} \sin\left(\frac{\pi j}{2}\right) \cos\left(\frac{\pi j x}{L}\right)    e^{-\alpha \left(\dfrac{ \pi j}{L} \right)^2 t}.
    \label{eq:exact-solution-heat}
\end{equation}

\begin{figure}[htpb]
	\centering
	\includegraphics[width=0.7\textwidth]{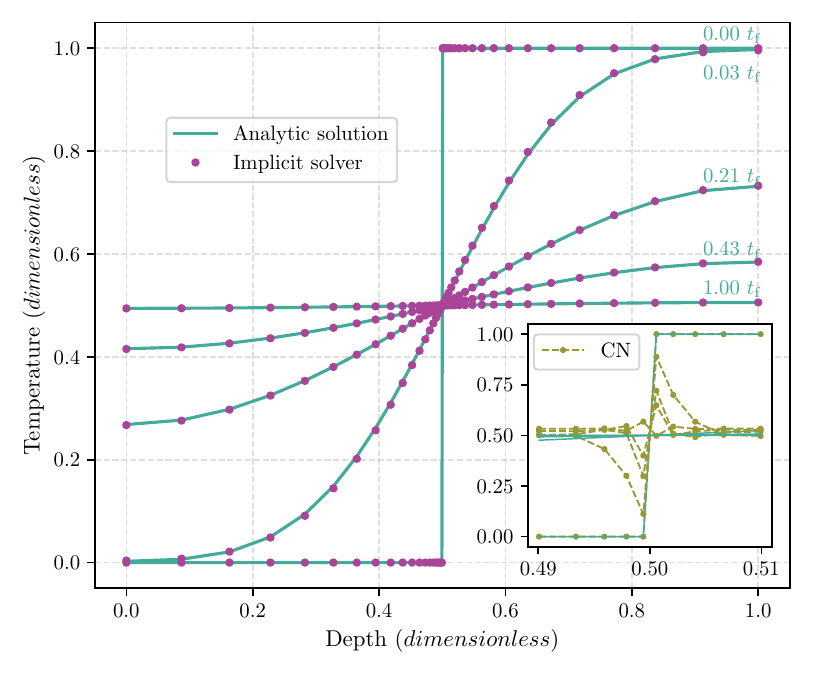}
	\caption{Analytical validation for a stiff initial condition and homogeneous thermal properties on an irregular grid for different timesteps. Here $t_{\mathrm{f}} = n_t \Delta t$. Despite the discontinuity at $x = 0.5 $, the fully implicit solver MultIHeaTS can compute the evolution of temperature with a very close match (maximum error $e_+< 0.5\%$) to the analytic solution. (\textit{Bottom Right}) Zoom on the spurious oscillations of the Crank-Nicolson solver at the location of the initial discontinuity. }
	\label{fig:valid_ana}
\end{figure}

To validate our model, we need to compare the computed temperatures with the analytical solutions for the same set of thermal parameters. This is done by computing the error, defined as the absolute value of difference between the temperature produced by the numerical model $T$ and the reference temperature $T_{\mathrm{R}}$. The maximum error $e_+$ is expressed as : 
\begin{align}
    e_+ = \max_{n, i}{ \left| \cfrac{ T^i_n - T_{\mathrm{R}, {n}}^i }{\langle T \rangle } \right|}
\end{align}
where $\langle T \rangle$ is the mean temperature and the average error $\overline{e}$ as 
\begin{equation}
    \overline{e} = \frac{1}{n_x n_t} \sum_{n, i} \left| \cfrac{ T^i_n - T_{\mathrm{R}, {n}}^i}{\langle T \rangle } \right|
\end{equation}
with $T_{\mathrm{R}, {n}}^i$ the reference temperature obtained from the analytical Equation \eqref{eq:exact-solution-heat} at location $x_n$ and time $t^i$.
For proper validation, both the numerical solver and the reference solution need to be calculated with exactly the same thermal properties and conditions (see Figure \ref{fig:valid_ana}). 
To showcase the advantages of using a solver capable of handling irregular grids, we specifically compute the temperature on an uneven spatial grid, strategically denser near the temperature discontinuity $\mu = L/2$. The grid spacing is determined by the following expression:
\begin{equation}
\begin{cases}
        x_0 = 0 \\
        x_n =  x_{n-1} + \dfrac{g(x'_n)}{\sum_1^{n_x-1} g(x'_k)}  L, \, \forall n \in \{1, \: \dotsc \: , n_x-1 \} 
\end{cases}
\end{equation}
where $g$ is normalized sigmoid function defined by
\begin{equation}
    g \colon x' \longmapsto 2 \left( 1+ e^{  \left(1- \frac{|x'-\mu|}{\mu}\right) /d  } \right)^{-1},
\end{equation}
where $d$ is a parameter controlling the streepness of the sigmoid and $x'_n = n/(n_x-1) \times L$. For a set of $n_x= 40$ grid points defining the box of length $L=\SI{1}{m}$ of constant diffusivity $\alpha = k / (\rho c_\mathrm{p}) \approx \SI{0.55}{m^2.s^{-1}}$ and $n_t=700$ iterations of timestep $\Delta t = \SI{2.3}{ms}$, the maximum error $e_+$ between the analytical solution and the fully implicit solver is less than 0.5\% and the mean error $\overline{e}$ is under 0.02\%.  
Results show a very close fit between the numerical and analytic solutions which proves that the MultIHeaTS model works well for homogeneous media.

The same test was performed using the Crank-Nicolson solver \citep{Schorghofer2022} to evaluate its ability to handle stiff initial conditions.
Using the same parameters as previously, the Crank-Nicolson solver produced spurious oscillations that were located around the initial discontinuity position (Figure \ref{fig:valid_ana}, \textit{Bottom Right}). 
The maximum error $e_+$ between the analytical solution and the Crank-Nicolson solver reaches up to 38\% and the mean error $\overline{e}$ averages 3\%. 
Although these oscillations eventually slowly disappear with time \citep{Oesterby2003}, it demonstrates that despite being the most accurate finite difference scheme, the Crank-Nicolson solver is less reliable in handling stiff initial conditions \citep{Press1992, Oesterby2003, Langtangen2017}, such as when two materials with different temperatures come into contact.
Such scenarios could arise in planetary science, such as when hot lava interacts with the Earth's surface, cryolava deposits on the icy surface of Europa, or CO$_2$ precipitates on the regolith of Mars.

%\textcolor{red}{One drawback of CN is that it responds to jump discontinuities in the initial conditions with oscillations which are weakly damped and therefore may persist for a long time.  }

\subsection{Comparison with Spencer's explicit Thermal Model}

\subsubsection{Validation by Numerical Comparison}

In more complex cases, in particular when the thermal conductivity vary with depth, the heat equation becomes too complex to solve analytically \citep{Jaeger1950}. 
In these cases, validation of the model is typically performed through comparison with experimental data and/or other well-established numerical algorithms.
Since our target application is the study of realistic planetary surface conditions, we decided to compare our method with Spencer's explicit scheme \citep{Spencer1989}, which is a commonly used algorithm in planetary science. 
Spencer's algorithm was implemented in IDL and can be obtained from the author's personal website\footnote{\url{https://www.boulder.swri.edu/~spencer/thermprojrs/}}.

%We used the same thermal properties, boundary conditions and numerical parameters (see Table. \ref{tab:params}), including the solar flux  $F_{\mathrm{solar}}$.

\begin{table}[htbp]
    \centering
    \begin{tabular}{ll}
    \toprule
    Parameters & Value \\ 
    \midrule
    Distance to Sun $d$ &  9.51 AU  \\
    Emissivity $\epsilon$ &  1  \\
    Albedo $A$ & 0.015  \\
    Grid points $n_x$ & 100  \\
    Max Depth $L$   & 2 m    \\
    Initial Temperature $T$ &  90 K  \\
    Solar Period $P$  & 79.3 days   \\
    Number of Periods & 5   \\
    Number of iterations $n_t$ &  50000   \\
    Time Step $\Delta t$ &  685 s   \\
    \bottomrule
    \end{tabular}
\caption{Physical and numerical parameters used for comparing Spencer algorithm with MultIHeaTS. While closely related to the Iapetus case, this dataset serves more as a reference scenario.}
\label{tab:params}
\end{table}

\begin{table}[htbp]
    \centering
    \begin{tabular}{ l l c c l } 
     \toprule
     Thermal Properties & Unit & Top value & Bottom value & Interface \\ 
     \midrule
     Density $\rho$ & kg.m$^{-3}$ & 800 & 2000 & 25 cm \\ 
    Heat Capacity $c_{\mathrm{p}}$ & J.kg$^{-1}$.K$^{-1}$ & 600 & 1800 & 50 cm \\ 
    Inertia $\Gamma_{\mathrm{th}}$ & J.m$^{-2}$.K$^{-1}$.s$^{-1/2}$ & 200 & 20 & 100 cm \\ 
     \bottomrule
    \end{tabular}
\caption{Thermal properties of a bilayered surfaced used for comparison of Spencer's explicit algorithm with MultIHeaTS. Although some of these values are close to what could be found on realistic icy surfaces, they were varied smoothly over large scales for validation purposes. }
\label{tab:thermal_ppties}
\end{table}

\begin{figure}[htpb]
	\centering
	\includegraphics{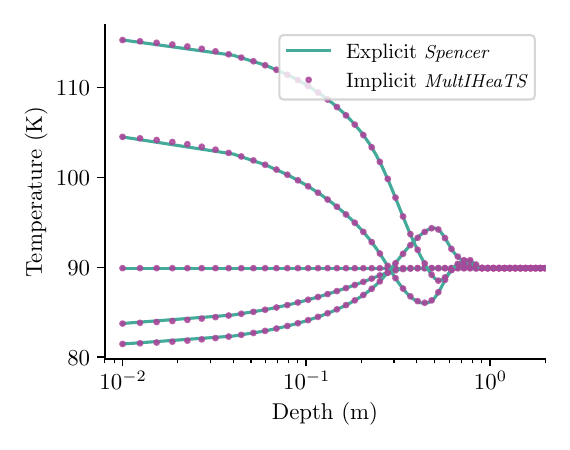}
	\caption{Validation of MultIHeaTS against Spencer's explicit algorithm. Temperature profiles for different times are plotted simultaneously $i = 0$, $200$, $14050$, $34959$, $49000$. MultIHeaTS is as accurate as Spencer's explicit solver with the advantage of stability. 
    Note the uneven grid spacing of MultiHeaTS, denser at the surface, shown by the logarithmic horizontal axis.
    }
    \label{fig:val_spencer}
\end{figure}

Both algorithms were run with meticulous attention with the exact same thermal properties and numerical parameters that can be found in Table \ref{tab:params}, along with identical solar flux at the surface given by Equation \ref{eq:slr_flux_model}, and a zero bottom heat flux.

The computation presented here were computed on an irregular grid given by the relation of recurrence:
\begin{equation}
   \forall n \in \{0, \: \dotsc \: , n_x-1 \} , \,  x_n = \left( \frac{n}{n_x -1} \right)^{p_{\mathrm{ow}}} L
   \label{eq:grid}
\end{equation}
where the exponent $p_{\mathrm{ow}}=4$ and $L$ is the bottom layer's depth. 
The characteristic depth of penetration of oscillating temperature wave with period $p$, is given by the thermal skin depth $\delta = \sqrt{\alpha p / \pi}$.
Such grid was chosen to increase spatial step with depth according to a power law, reflecting the exponential decrease in temperature variations and reduced spatial precision needed after the diurnal skin depth $\delta_{\mathrm{d}}$.
For the sake of validation, values of density, heat capacity and conductivity are varied through the depth by the largest scales allowed by Spencer's explicit scheme stability criteria (see Table \ref{tab:thermal_ppties}).
%The MultIHeaTS solver completed these 50 000 iterations of a 100-point grid in approximately one second on an Intel CPU i7-10750, showcasing its efficient computational performance.

%on a $\SI{2}{m}$ thick grid consisting of $n_x=100$ points for $n_t = 5 \times 10^4 $ iterations, 

Overall, the results, presented in Figure \ref{fig:val_spencer},  show a very good similarity between the temperature profiles produced by the explicit and fully implicit schemes, with negligible differences between the two methods.
The maximum error between MultIHeaTS and the Spencer's solver is $e_+=0.68\%$ and the mean error is $\overline{e}=0.052\%$.
%\textcolor{red}{Errors are mostly concentred near the surface (in opposition to CN where they are at the highest temperature gradient, near the bilayer interface.}
This comparison demonstrates that the MultIHeaTS solver can accurately compute temperature profiles for planetary surfaces, without the limitations of conditional stability that can be encountered with explicit methods.

\subsubsection{Computational Efficiency of Implicit Solvers}
\label{sec:efficiency}

\begin{figure}[htpb]
	\centering
	\includegraphics[width=\textwidth]{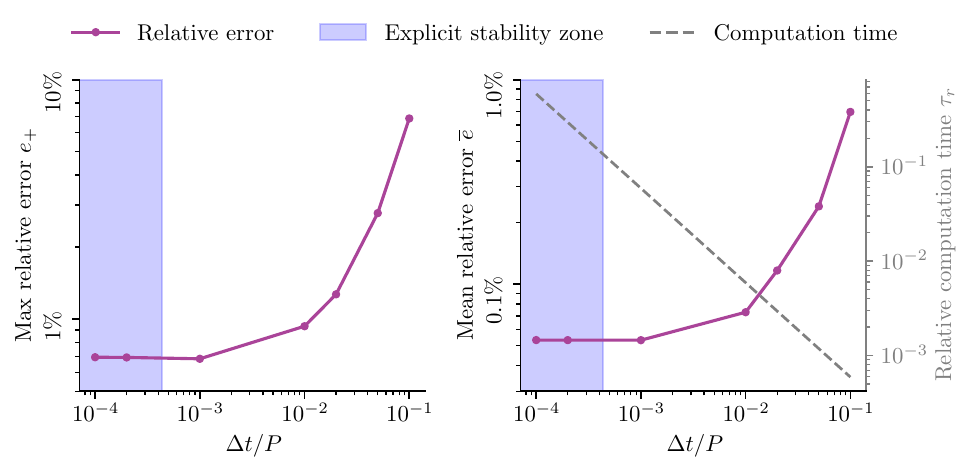}
 \caption{
    Relative errors produced by the fully implicit scheme when increasing the timestep $\Delta t$ (log-scales). The reference is Spencer's explicit thermal model computed with $10^4$ points per day.  Timestep $\Delta t$ is represented as a fraction of the surface flux’ period $P$. 
    The stability zone of the explicit scheme $F < 1/2$ is represented in
    light blue. (\textit{Left}) Maximum error $e_+$. (\textit{Right}) Average error $\overline{e}$ and relative computation time $\tau_r$. The computational advantage of the fully implicit solver enables much faster temperature calculations.
    }
    \label{fig:speed_spencer}
\end{figure}

While both the explicit and fully implicit schemes can compute the same temperature, in this section we focus on comparing the computational speed of these solvers.
On a 6 cores Intel i7-10750H CPU at 3.6 GHz, using the parameters shown in Tables \ref{tab:params} and \ref{tab:thermal_ppties}, Spencer's thermal model is capable of performing 27,000 iterations per second. In contrast, MultIHeaTS on the same hardware can achieve up to 47,000 iterations in the same time, resulting already in a 60\% faster computation.
However, the primary advantage of using an unconditionally stable solver lies in its ability to remove restrictions on the parameter space. Unlike the Spencer's explicit model, which is limited by the stability criteria defined in Equation \ref{eq:stability}, MultIHeaTS can handle arbitrarily large timesteps.

To demonstrate this, Figure \ref{fig:speed_spencer} depicts the temperature difference between our fully implicit scheme and the Spencer scheme as a function of the timestep. We observe that as the timestep increases, the differences gradually increase as well. Nonetheless, even with a timestep 100 times larger than the reference ($\Delta t / P \sim 10^{-2}$), the maximum relative error $e_+$ remains below 1\% and the mean relative error $\overline{e}$ remains below 0.1\%.
This indicates that we can compute the same temperature with significantly larger timesteps while maintaining a reasonably low error.
It is important to note that comparing the Crank-Nicolson scheme, a second-order method in time, to the first-order Spencer scheme would not yield meaningful results. However, due to its second-order precision, the semi-implicit scheme will show higher accuracy in terms of mean error compared to the other two schemes when compared to the ground truth temperature.

We conducted a comparison of the computational speed required by both solvers to calculate the final temperature after a given time $t_\mathrm{f} = 5 P$. To quantify this, we introduced the relative computational time $\tau_r$, defined as:

\begin{equation}
\tau_r = \frac{\tau_{\mathrm{implicit}}}{\tau_{\mathrm{spencer}}}
\end{equation}
where $\tau_{\mathrm{spencer}}$ represents the computational time required for the reference Spencer model to compute the final temperature, while $\tau_{\mathrm{implicit}}$ represents the time taken by MultIHeaTS to compute the same final temperature.

Figure \ref{fig:speed_spencer} (Right) illustrates the computational advantage of the implicit schemes. By requiring fewer iterations, the total computational time of the fully implicit solver $\tau_{\mathrm{implicit}}$ is significantly smaller than that of the explicit scheme $\tau_{\mathrm{spencer}}$.
Consequently, our MultIHeaTS solver can compute the same temperature for a given time up to 100 times faster than the explicit method while staying accurate. This is particularly advantageous for simulating processes on large timescales. For instance, if we aim to determine the temperature after one million years to study the thermal evolution of the icy crust of a Galilean moon, a computation that would take a year on the explicit solver could be reduced to less than 4 days using MultIHeaTS.

\section{Applications}

\subsection{Bilayer Thermal Signature for Remote Sensing}
\label{sec:application}

Surface temperature variations are crucial measurements obtained by space probes for inferring the thermal properties of planetary surfaces, using measurement of brightness temperature in infrared wavelength such as THEMIS on Mars \citep{Christensen2004} or CIRS for the Saturn system \citep{Jennings2017}. The same principle also applies on radar frequencies, such as Cassini RADAR passive-mode for the Saturn System \citep{Elachi2004}. 
Therefore, accurately simulating the temperature evolution of complex profiles becomes essential for comparing with measurements and extracting valuable properties through inverse problems.
Here, we applied our thermal model to Europa's surface at latitude and longitude (0\textdegree, 180\textdegree) to investigate the differences in surface temperatures without eclipses and  explore the thermal signatures associated with four distinct profiles of thermal properties.

First, we assume two types of homogeneous icy surfaces. The first one, called material ``H'' (for high thermal inertia), is a porous but bulkier water ice a heat capacity of $c_{\mathrm{p}}=\SI{839}{J.kg^{-1}.K^{-1}}$ \citep{Klinger1981}, and a thermal inertia of $\Gamma_{\mathrm{H}}=\SI{196}{SI}$. 
This ``H'' material has significantly lower thermal inertia than bulk ice at these temperatures \citep{Klinger1981}, yet it is termed "high" because its values exceed those derived from brightness temperature measurements on Europa \citep{Rathbun2010, Trumbo2018}.
The second one, called material ``L'' (for low thermal inertia), is a icy regolith similar to the highly porous lunar type regolith. This leads to modified values for density $\rho =\rho_{\mathrm{0}} (1-\phi)$ and heat conductivity $k = k_{\mathrm{0}}(1-\phi)$. In the case of the small grain deposit, the material L can have very high porosity. We chose a porosity of $\phi \sim \SI{85}{\%}$, resulting in low thermal inertia of $\Gamma_{\mathrm{L}}=\SI{28}{SI}$. 

The third test case is a bilayered profile made of material L over H, called bilayer$_{\mathrm{LH}}$ which can represent a layer of small grains  deposited from a plume on top of a bulk icy surface. 
The fourth case is a bilayered profile of material H over L, called bilayer$_{\mathrm{HL}}$ which can represent cryolava deposited over a highly porous icy regolith-type, it leads to a situation where a material of high thermal inertia H can overlay a material of low thermal inertia L.

The main question is whether it is possible to distinguish these four situations from spaceborne measurement that can only access surface temperature. If such a distinction is achievable, what recommendations should be considered for mission planning to effectively distinguish these situations?

\begin{table}[htbp]
    \centering
    \begin{tabular}{ll}
    \toprule
    Parameters & Value  \\ 
    \midrule
    Interface HL $d_{\mathrm{HL}}$ & \SI{5.00}{cm} \\
    Interface LH $d_{\mathrm{LH}}$ & \SI{1.00}{cm} \\
    Diurnal skin depth  of H  $\delta_{\mathrm{dH}}$ &  \SI{7.97}{cm} \\
    Diurnal skin depth  of L  $\delta_{\mathrm{dL}}$ &  \SI{1.13}{cm} \\
    Low thermal inertia  $\Gamma_{\mathrm{L}}$ &  \SI{27.8}{SI}  \\
    High thermal inertia  $\Gamma_{\mathrm{H}}$ &  \SI{196}{SI}  \\
    Emissivity $\epsilon$ &  0.94  \\
    Bond albedo $A$ & 0.6  \\
    Grid points $n_x$ & 100  \\
    Max depth $L$   & 6 m    \\
    Initial temperature $T$ &  105 K  \\
    Number of Europa days & 5   \\
    Number of iterations $n_t$ &  300   \\
    Time step $\Delta t$ &  5000 s   \\
    \bottomrule
    \end{tabular}
\caption{Physics and numerical parameters used for the thermal signature application, based on the Europa case. The diurnal period is $p_{\mathrm{E}} = \SI{3.55}{days}$ and  the SI unit of thermal inertia is $\unit{J.m^{-2}.K^{-1}.s^{-1/2}}$}
\label{tab:signature}
\end{table}

To answer this question, we compare different types of profiles made of the aforementioned materials. Importantly, the interface depth of both bilayer profiles was deliberately selected to be thinner than the thermal diurnal skin depth $\delta_{\mathrm{d}}$ (see Table \ref{tab:signature}). This choice is crucial because when the interface is significantly deeper than the skin depth, then the surface temperatures are predominantly influenced by the upper layer  and indistinguishable from the case of constant properties. Depending on the interface depth, between 20 to 30 grid points are used to describe the top layer.

To obtain accurate simulations, we incorporate solar flux data from NASA's SPICE Toolkit \citep{Acton1996, Acton2018} arbitrarily chosen at UTC 23 August 1997, which provides precise information on the distance to the sun and solar incidence at the specific point on Europa's surface. These parameters are integrated instead of the simplified solar flux from Equation \ref{eq:slr_flux_model}, allowing us to run realistic simulations over a span of 5 Europa days.

A similar study conducted by \cite{Putzig2007} on Mars using THEMIS presented results in terms of apparent thermal inertia instead of surface temperature. They varied the upper layer thickness within the range of $\delta_{\mathrm{d}}/512$ to $\delta_{\mathrm{d}}$, where $\delta_{\mathrm{d}}$ represents the skin depth for Mars' dust (approximately $\SI{21}{cm}$).

\begin{figure}[htpb]
	\centering
	\includegraphics[width=0.7\textwidth]{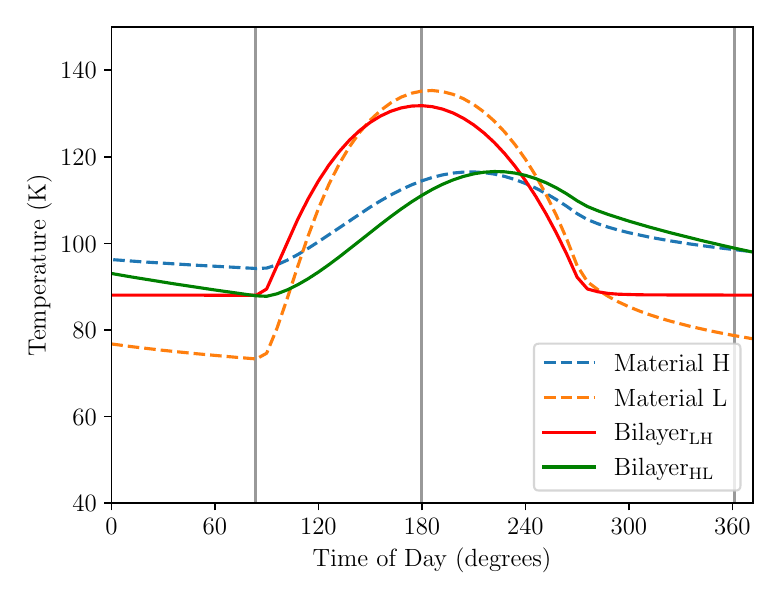}
	\caption{Daily surface temperatures on Europa for four different ground profiles at Lat/Long (0, 180). Dashed lines show monolayered profiles and solid lines show bilayered profiles. No homogeneous ground profile could match the thermal signature of bilayered ones. The grey vertical lines indicate the recommended measurement time for a space probe to discern between a homogeneous surface and a bilayered one.}
	\label{fig:surf_temp}
\end{figure}

The results in Figure \ref{fig:surf_temp} show the expected behavior in the case of homogeneous thermal inertia.
In the case of low thermal inertia (L), the peak with highest temperature is close to local noon (Time of Day $180^{\circ}$), the maximum surface temperature is close to the equilibrium temperature and night time temperature are relatively cold and cooling significantly during the night.
On the contrary, for high thermal inertia (H), the peak with highest temperature is shifted toward the afternoon, the maximum surface temperature is lower and night time temperatures are warmer \citep{Putzig2007}.
The results shown in Figure \ref{fig:surf_temp} also reveal the particular trends of bilayered profiles. In the particular case of an interface close to the thermal diurnal skin depth, the expected trends are mixed. 
The bilayer$_{\mathrm{LH}}$ has a high centered day-side temperature bell shape and almost constant night temperatures, consistent with the observations from bilayer thermal models by \citet{Squyres1980}. In contrast, the bilayer$_{\mathrm{HL}}$ has a shifted day-side temperature peak and strong cooling at the end of the night.
Over the range of inertia bounded by $\Gamma_{\mathrm{L}}$ and $\Gamma_{\mathrm{H}}$, no homogeneous ground profile could match the thermal signature of bilayered ones. We thus recommend to observe the same surface at three strategic local times as shown by the grey vertical lines in Figure \ref{fig:surf_temp}.
Measuring the surface at the morning sunrise, at noon and at  midnight, would help a distinguish a bilayered surface with an interface close to the diurnal skin depth to a homogeneous surface.

This strategy would be interesting for the preparation of the upcoming missions, such as JUICE \citep{Grasset2013} and Europa Clipper \citep{Phillips2014}. Solvers like MultIHeaTS, that can accommodate any observations set (whatever their observation time) with depth-dependant properties, will be highly valuable in enhancing our understanding of Europa's surface profiles. 
Whether the surface exhibits, for example, homogeneous porosity of ice grains or a depth-dependent variation, distinct brightness temperature patterns would be observed and measured. 
In addition to the surface temperature profiles, radiometer or sub-millimeter instruments can probe brightness temperature at different wavelengths to distinguish materials at various depths.
By comparing the solver's output with actual measurements, valuable information regarding the surface properties can be retrieved.

\subsection{Million Year Simulation of Europa}

\subsubsection{Orbit and Solar flux}

\begin{figure}
    \centering
    \includegraphics[width=0.8\textwidth]{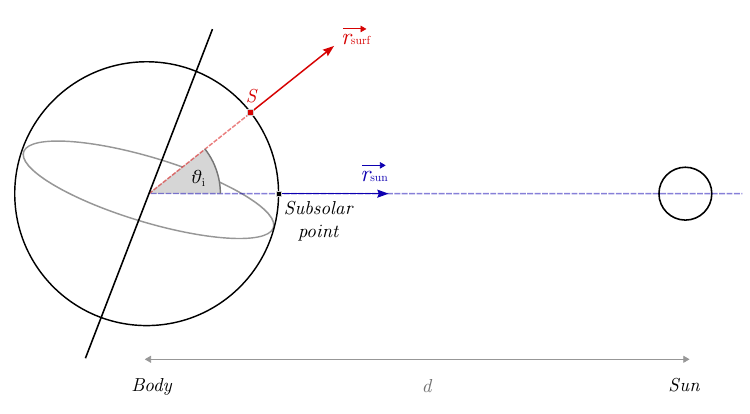}
    \caption{Schematic representation of the target body and Sun system.  The solar incidence angle $\theta_i$ is defined as the angle between the normal vector at the target surface of the body $r_{\mathrm{surf}}$ and the normal vector at the sub-solar point $r_{\mathrm{sun}}$.}
    \label{fig:solar-inc}
\end{figure}

To apply the model to Europa over a million-year timescale, we must account for variations in solar flux during this period.
During periods of day, for a given time, latitude $\lambda$ and longitude $\psi$, the solar flux is given by \citet{Spencer1989}:
\begin{equation}
	F_{\mathrm{solar}}(t, \lambda, \psi) = 
			\left(1 - A(\lambda, \psi) \right) \dfrac{G_{\mathrm{SC}}}{d(t)^2} \cos(\theta_i(t,\lambda, \psi))
 \label{eq:slr_flux_model_accurate}
\end{equation}
where $A$ is the surface bond albedo, $G_{\mathrm{SC}}$ the solar constant, $\theta_i$ the solar incidence angle and $d$ the distance to the Sun in AU (see Figure \ref{fig:solar-inc}).

\paragraph{Distance to the Sun and orbital parameters}
To determine the distance to the Sun, we initially need to obtain Europa's position in its orbit around Jupiter, which gradually varies over a million year. Given the negligible difference in distance between Europa and its planet, compared to Jupiter's distance from the Sun, we can directly calculate the flux received by Jupiter, which deviates by less than $0.02\%$ from that received by Europa. 

Describing a body's motion around the Sun can be done using its orbital elements, such as the semi-major axis $a$, eccentricity $e$, perihelion longitude $\omega$, inclination $I$, ascending node longitude $\Omega$, and mean longitude $\Psi$. 
For a period of million years, some orbital elements remain relatively constant. 
For instance, the semi-major axis remains stable at $a = \SI{5.204}{AU}$ in the secular regime \citep{Laskar2008}.
Similarly, due to the near-flat inclination of Europa's orbit to Jupiter's equator plane and low Europa's obliquity \citep{Bills2009}, we fix theses values to 0. 
In reality, the sub-solar latitude on Europa varies by $\pm \ang{3.7}$ \citep{Ashkenazy2019}, which would need to be considered for high-latitude computations but is beyond the scope of this paper.
Moreover, as the perihelion longitude $\omega$ and ascending node longitude $\Omega$ do not influence solar flux, we assigned them arbitrarily, to their values at J2000, using VSOP2013 \citep{Simon2013}.
%: $\omega = 0.2715 $ and $\Omega = ?$.

However, the eccentricity parameter can fluctuate over a million-year timescale, leading to changes in solar flux which necessitates consideration \citep{Laskar2008}.
For short simulations, precise estimates of the eccentricity could be obtained using SPICE kernels, but these are only available for Europa for up to a thousand years, not a million. 
In order to obtain the periodicity of eccentricity, we used the studies from  \cite{Laskar2003, Laskar2008}, which conducted frequency analysis of Jupiter's motion over $\SI{50}{My}$. 
Their results show that the quasiperiodic approximation of eccentricity can be expressed as the sum of five harmonics:
\begin{equation}
e = e_0 + \sum_{i=1}^{5} e_i \cos \left( \nu_i t + \upvartheta_i \right)
\label{eq:ecc}
\end{equation}
where the values of $e_i$, $\nu_i$ an $\upvartheta_i$ are found in Table 7 of \cite{Laskar2008}.
These five harmonics have periods spanning from 27,000 years to 1.1 million years, with eccentricity values fluctuating  over the course of our simulation from $e = 0.027$ to $e=0.061$. 
At maximum eccentricity ($e = 0.061$), Jupiter’s heliocentric distance ranges from 4.88 to 5.52 AU within an orbit, while at minimum eccentricity ($e = 0.027$), it varies between 5.04 to 5.35 AU. Based on Equation \ref{eq:slr_flux_model_accurate}, this leads, for example, to a variation of the maximum solar flux at the equator of up to $\sim 7 \%$ at the same solar longitude.

Finally, the mean longitude changes periodically over one orbit, and is updated over time using the equation:
\begin{equation}
   \Psi(t) \equiv  \Psi(t_{J}) + \frac{2 \pi }{p_{\mathrm{J}}} (t - t_{J}) \;(\bmod{\; 2 \pi}) 
\end{equation}
where $p_{\mathrm{J}}$ is the orbital period of Jupiter and $t_{J}$ is the reference time at J2000.

At each iteration, the six orbital elements are computed and passed to the \mbox{ELLXYZ} subroutine from VSOP2013 \citep{Simon2013}, which calculates the planetary rectangular coordinates.
Then, these coordinates are used to determine the distance from Europa to the Sun for each iteration, which is essential for computing the solar flux.

\paragraph{Solar Incidence}
Now to compute the solar incidence of a particular point on Europa's surface at a given latitude $\lambda$ and longitude $\psi$, we express the surface normal and the sub-solar point surface normal unit vector as:
\begin{equation}
  \vec{r}_{\mathrm{surf}}=
  \begin{pmatrix}
  \cos(\lambda) \cos(\psi)\\
  \cos(\lambda) \sin(\psi)\\
  \sin(\lambda) \\
  \end{pmatrix} 
  , \quad
  \vec{r}_{\mathrm{sun}}=
  \begin{pmatrix}
   \cos(\psi_{\mathrm{sub}})\\
   \sin(\psi_{\mathrm{sub}})\\
  0 \\
  \end{pmatrix} 
\end{equation}
with the sub-solar point longitude being updated as
\begin{equation}
   \psi_{\mathrm{sub}}(t) \equiv  \psi_{\mathrm{sub}}(0) + \frac{2 \pi }{p_{\mathrm{E}}} t \;(\bmod{\; 2 \pi}) \\
\end{equation}
where $p_{\mathrm{E}}$ is the orbital period of Europa and we arbitrarily take $\psi_{\mathrm{sub}}(0)=0$. In absence of other information, we assumed that the rotation period of Europa remains constant.
%\textcolor{red}{This makes the assumption that each day is the same and it this does not change through one million year. Citer pourquoi est ce qu'on peut faire ça ?}
The solar incidence angle $\theta_i$ is then given by the scalar product between the surface's normal and the sub-solar point normal:
\begin{equation}
    \theta_i = \arccos \left( \vec{r}_{\mathrm{sun}} \cdot \vec{r}_{\mathrm{surf}} \right)
\end{equation}
Then for periods of day, when $\theta_i  \text{ in } \left[-\frac{\pi}{2}, \frac{\pi}{2}\right]$, the solar flux is computed using Equation \ref{eq:slr_flux_model_accurate} and for periods of night, when $\theta_i  \text{ in } \left]\frac{\pi}{2}, \frac{3\pi}{2}\right[$, the solar flux is zero $F_{\mathrm{solar}} = 0$.
At each iteration of the model, the solar flux is computed for current time and injected at the top boundary condition of the thermal solver to compute the heat transfer.

 \subsubsection{Thermal Properties}

Europa surface and subsurface is made mainly of water ice \citep{CruzMermy2023}. 
There are number of evidence showing that its icy surface is porous.
For instance using spaceborne thermal infrared instrument, the thermal inertia is in the range $\Gamma = 40-\SI{150}{W.m^{-2}.K^{-1}.s^{-1/2}}$ \citep{Spencer1999, Rathbun2010}. 
Using Earth-based observations, another team obtained a thermal inertia of $\Gamma = \SI{95}{W.m^{-2}.K^{-1}.s^{-1/2}}$ \citep{Trumbo2018}.
Theses values are order of magnitude inferior to the thermal inertia of solid ice at these range of temperature $\Gamma_{\mathrm{ice}} \approx \SI{2600}{W.m^{-2}.K^{-1}.s^{-1/2}}$ \citep{Klinger1981}.
Such difference can be explained by adding porosity into the ice, significantly lowering its thermal inertia \citep{Spencer1999}.
The thermal inertia depends on thermal parameters according to the expression $\Gamma = \sqrt{k\rho c_p}$.
Although porosity doesn't directly influence the specific heat capacity, it does alter ice density and conductivity.

\paragraph{Ice Density and Porosity}

To obtain the porous ice density, we must evaluate if the porosity can vary due to compaction at the depths affected by temperature variations caused by the periodic oscillations of solar flux input at the surface.
Various compaction mechanisms can intervene (e.g., sputtering or sintering induced); however, we focus solely on gravity-induced compaction for this analysis. Following, \cite{Mergny2024c}, the expression for density as a function of depth is
\begin{equation}
    \rho(x) = \rho_{\mathrm{0}} \left(1 + \frac{\phi(0)}{1-\phi(0)} \exp\left(-\dfrac{x}{H} \right)  \right)^{-1}
    \label{eq:density_wilson}
\end{equation}
where $\rho_{\mathrm{0}}$ is the bulk ice density, $\phi(0)$ the surface porosity and $H$ the characteristic lengthscale of compaction.

The diurnal and seasonal thermal skin depths on Europa are much shallower than the gravity-driven compaction length scale $H$, which is on the order of hundreds of meters \citep{Mergny2024c}.
However eccentricity variations happen on large timescales (Equation \ref{eq:ecc}) and are highly dominated by a $\sim \SI{54}{kyrs}$ period harmonic, which lead to a "geological" thermal skin depth of up to $\SI{130}{m}$.
Our simulations are computed up to depth $L = \SI{50}{m}$, which results in small variations, up to $\SI{1}{K}$, of the bottom temperature within geological timescales.
Extending to depths of hundreds of meters would 1) increase the computation cost 2) increase complexity due to density being affected by gravity at these depths. While feasible, these computations are beyond our focus ; MultiHeaTS is part of a multiphysics model, LunaIcy, simulating near-surface interactions on the ice microstructure \citep{Mergny2024i}, with processes like ice grain sintering only having noticeable effect within the first few meters.
Other compaction processes, such as those induced by micrometeorite impacts will influence the density of the near surface, but these are not considered in this study and remain to be explored in future work.
Therefore, at the depths relevant to our thermal analysis, $x \ll H$, the porosity is assumed to remain constant, resulting in the simplified expression for the density of porous ice
\begin{equation}
    \rho(\phi) = \rho_{\mathrm{0}} (1- \phi).
\end{equation}

\paragraph{Porous Ice Conductivity}

Number of studies have work on formulating the conductivity of  compacted spheres.
One common approach describes thermal conductivity as a function of the porosity $\phi$ as $k(\phi) = k_{b} (1-\phi)$ \citep{Ferrari2005}. 
However, to reach the low thermal inertia of Europa reported in the literature \citep{Spencer1999}, such approach  would require a porosity of $\phi=0.99$, which seems unlikely.
One important aspect that this first formulation does not take into account is the quality of contact between grains.
Considering that ice conducts heat much better than radiation in the pores, accounting for the contact area between grains becomes crucial in porous materials.  The pores are at near-vacuum so heat transfer across pores is likely predominantly by radiation.
This concept was explored in an early study by \cite{Adams1993}, where they developed a physical model for snow conductivity of packed spheres, distinguishing between ice conduction, conduction in the pore space, and vapor transport. Subsequent papers, such as the SNOWPACK model for avalanche warning \citep{Lehning2002b}, adopted the same formulation with refinements.
Models like \citet{Gusarov2003} have also addressed the conductivity of packed spheres, particularly in cases where there is a limited contact region between grains.
Some difficulties arise from these formulations as they rely on the coordination number, which is often calculated through empirical means based on snow data on Earth. 
For these reasons, these models are appropriate for highly porous material with small regions of contact, but for low porosity, they lead to porous conductivity higher than the bulk ice conductivity which is not realistic.
%In the context of Europa's conditions, the contributions of air and radiation conductivity, as well as vapor transport, can be considered negligible compared to ice conductivity

Another method, is to use the Hertzian theory to determine the efficiency of  heat exchanged between grains \citep{Gundlach2012}.
This effect is taken into account by the introduction of the
so-called Hertz factor, which accounts for the reduction in effective cross-section area of the porous material due to its porosity. 
While numerous studies have explore in details how to express the Hertz factor as a function of the temperature and grain size, here we adopt the definition of the Hertz factor as the ratio between the radius of contact $r_{\mathrm{b}}$ and the grain radius $r_{\mathrm{g}}$.
When two grains come into contact, they deform elastically and the initial radius of contact depends on the forces applied on the grains.
For ice grains, the adhesion stage occurs due to the Van der Waals interaction between particles, leading to the expression of the bond radius \citep{Molaro2019}:
\begin{equation}
    r_{\mathrm{b}} \approx \left( \dfrac{\gamma r_{\mathrm{g}}^2}{ 10 \mu} \right)^{1/3}
    \label{eq:bondVDW}
\end{equation}
where $\gamma$ is surface tension of water ice and $\mu$ its shear modulus.

Following \cite{Ferrari2016}, the expression of the porous conductivity can then be generalized to
\begin{equation}
    k(\phi, r_{\mathrm{b}}, r_{\mathrm{g}}, T) = k_{\mathrm{0}}(T) S(\phi) \mathcal{H}(r_{\mathrm{b}}, r_{\mathrm{g}})
\end{equation}
where $k_{\mathrm{0}}$ is the conductivity of bulk ice for a given temperature, $\mathcal{H}$ is the Hertz factor and $S(\phi)$ is a factor accounting for the reduction of conductivity due to porosity with multiple formulations existing in the literature \citep{ Shoshany2002, Gusarov2003, Gundlach2012, Ferrari2016}.
In this study, to avoid empirical expressions, we favor a straightforward analytical form $S(\phi) = 1- \phi$ which lead to the expression of the porous conductivity
\begin{equation}
    k(\phi, r_{\mathrm{b}}, r_{\mathrm{g}}) = k_{\mathrm{0}}(T) \left(1-\phi \right) \frac{r_{\mathrm{b}}}{r_{\mathrm{g}}}.
    \label{eq:cond_porous}
\end{equation}
where the bulk ice conductivity for a given temperature is computed using the expression $k_{\mathrm{0}} = 567/ T$ \citep{Klinger1981}.
Updating the temperature-dependent conductivity requires to recalculate the tridiagonal matrix terms of Equation \ref{eq:an_terms}, which is computationally expensive. A performance test conducted for a grid of 100 points, reveals that this results in a slowdown of computation by a factor 2.
Therefore, for the current study  we maintain a constant bulk ice conductivity based on the surface equilibrium temperature.
While temperature variations on Europa do not have a significant impact on conductivity compared to porosity and grain contact changes, it can be valuable to include it in future work when the temperature is highly varying.
%The expression Equation \eqref{eq:cond_porous} used in this study, has been chosen for the simplicity of its analytical form.\
Also, while the radius of contact $r_{\mathrm{b}}$ and porosity $\phi$ are treated as free parameters, they inherently influence each other.
As the radius of contact approaches the grain radius, the pore space decreases, leading to a reduction in porosity towards zero. 

For this simulation, we used the following parameters: porosity of $\phi = 0.4$, grain size of $r_{\mathrm{g}} = \SI{100}{\micro\metre}$, and a temperature of $\SI{100}{\kelvin}$. Equation \eqref{eq:bondVDW} yields a contact area of $r_{\mathrm{b}} = \SI{0.39}{\micro\metre}$ and Equation \eqref{eq:cond_porous} results in a conductivity of $k = \SI{0.013}{W.m^{-1}.K^{-1}}$ for the porous medium.

\subsubsection{Anti-Jovian Hemisphere}

Using the MultIHeaTS thermal solver combined with our solar flux model and thermal properties, we can simulate heat transfer on Europa's icy surface for a million year in about four days of computation time. 
With this numerical approach, we have flexibility with the parameters used in the simulation, such as albedo, surface porosity, latitude, and longitude.

\begin{figure}[htpb]
	\centering
	\includegraphics{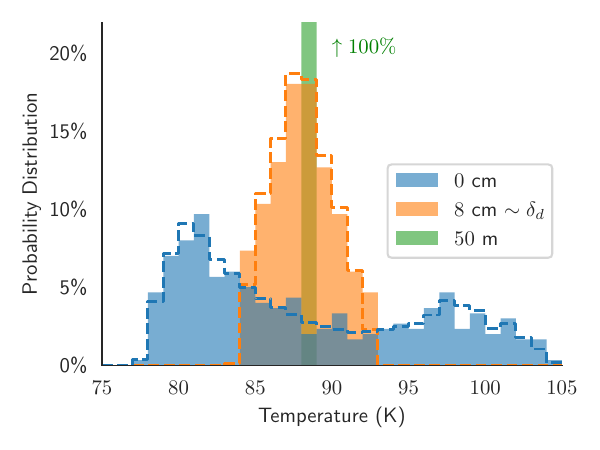}
	\caption{Temperature distribution on Europa at Lat/Long ($\ang{0}, \ang{180}$) with albedo $A=0.8$ for three strategic depths: at the top layer, close to the diurnal thermal skin depth, and at the bottom layer. The histogram shows the frequency of each temperature at each depth. Filled bars represent the one million year simulation, and dotted lines represent a single Jupiter year. }
	\label{fig:histo_temp}
\end{figure}

In this section we present the result of a simulation on the anti-Jovian hemisphere with latitude $\ang{0}$, longitude $\ang{180}$ and a porosity of $\phi = 0.4$. This leads to a thermal inertia of $\Gamma = \SI{83}{W.m^{-2}.K^{-1}.s^{-1/2}}$, coherent with the observations \citep{Spencer1989, Trumbo2018}, without needing excessive value of the porosity.
Here, the grid spacing is given by Equation \ref{eq:grid} with the exponent $p_{\mathrm{ow}}=5$ and maximum depth $L=\SI{50}{m}$.
This simulation was specifically computed for high albedo, here $A = 0.8$ and $\epsilon = 0.9$, so that temperatures remain low enough (less than 105 K) such that ice sintering is negligible during the million year timeframe.
The goal here is first to look at the temperature distribution without eclipses (see Fig. \ref{fig:histo_temp}), in a condition where the ice microstructure remains constant.
The joint paper \citep{Mergny2024i} expand upon this with low albedo simulations, by coupling the MultIHeaTS thermal solver to an ice sintering model that also changes the thermal properties.

The temperature distribution after one million year are plotted in Figure (\ref{fig:histo_temp}, \textit{filled bars}) for three strategic depths: at the top layer, close to the diurnal thermal skin depth and at the bottom layer.
Long integration time is require to converge towards a stable regime at depth. We have found that, after 20 orbits, the temperature at depth remains constant to seasonal variations, thus we exclude the initial $\sim$ 250 years from the temperature distribution.
Given the extensive computational load of simulating one million years, storing each of the three billion iterations is impractical due to size constraints. 
To address this, we opted for periodic saving of the temperature profile at asynchronous intervals, specifically at $p_{\mathrm{save}} = 10.1 \times p_{\mathrm{J}}$, ensuring that we capture all diurnal and orbital variations. Figure \ref{fig:histo_temp} illustrates the attenuation of the heat wave, noticeable around depths corresponding to the diurnal thermal skin depth, $\delta_{\mathrm{d}} = \SI{5.6}{cm}$.
While the bottom layer has a constant temperature, temperature variations are significantly higher near the surface.
We observe an asymmetric  temperature distribution for the top surface, with night-time temperatures predominantly colder than the equilibrium temperature ($\SI{89}{K}$).
Some of the surface temperatures exceed the equilibrium by more than 15 K, which are crucial for temperature-dependent processes such as sintering, as investigated in the joint paper \citep{Mergny2024i}.

Such distribution differs from one that would be acquired through a single orbit of Jupiter (Figure \ref{fig:histo_temp}, \textit{dotted lines}), due to the periodic changes of the orbital elements throughout a million year. 
While the temperature histograms have similar trends, some noticeable differences emerge, especially at high temperature ($\sim \SI{100}{K}$) which is relevant for precise estimation of temperature dependent processes.
The one million year temperature distribution can also be beneficial for one-way coupling models. For instance, if a certain temperature is reached with a probability $p(T)$ during a simulation lasting $t_{\mathrm{f}}$, it could be analog to a model maintaining a constant temperature $T$ during a time $t = p(T) \times t_{\mathrm{f}} $.

\subsubsection{Sub-Jovian Hemisphere}

In contrast to the anti-Jovian hemisphere, the sub-jovian hemisphere has multiple interactions with Jupiter that affects the diurnal flux reaching the surface. 

\paragraph{Jupiter's Emitted and Reflected Flux }

Jupiter emits longwave radiation that can affect Europa's surface temperature. Since Europa is tidally locked to Jupiter, the sub-Jovian hemisphere always absorbs this radiation.
Based on the emission temperature and distance to Europa, \citet{Ashkenazy2019} computed the radiation flux with a maximum at the equator of $\SI{0.056}{W.m^{-2}}$, very close to Europa's estimated internal heating.
This value is orders of magnitude smaller than the solar flux at Europa's heliocentric distance  $\sim \SI{50}{W.m^{-2}}$, and thus, like internal heating, it will be neglected in this study.

In addition to emitting radiation, Jupiter also reflects some of the solar flux it receives back to Europa.
The reflected flux from the sun by Jupiter, $F_{\mathrm{R}}$, can be modeled as
\begin{equation}
F_{\mathrm{R}} = A_{\mathrm{J}} \left( \dfrac{R_{\mathrm{J}}}{d_{\mathrm{E}}} \right)^2 \dfrac{G_{\mathrm{SC}}}{d^2}
\end{equation}
where $R_{\mathrm{J}} = \SI{70 000}{km}$ is Jupiter's radius, $A_{\mathrm{J}} = 0.34$ is Jupiter's Bond albedo, and $d_{\mathrm{E}} = \SI{671 000}{km}$ is the distance between Europa and Jupiter. This computation leads to a reflected flux value of $F_{\mathrm{R}} \approx \SI{0.18}{W.m^{-2}}$, which is still two orders of magnitude lower than the solar flux at Europa's heliocentric distance.
However, since this flux can reach the surface during nighttime and is about three times higher than the longwave radiation, it is not clear if it leads to significant temperature variations during the night.
Using Equation \ref{eq:energy-eq} at equilibrium and substituting the solar flux with $F_{\mathrm{R}}$, we find an equilibrium temperature of $\SI{36}{K}$, which is still much colder than Europa's nighttime temperatures (see Figure \ref{fig:histo_temp} for reference).
For context, in terms of the time of day, this flux is equivalent to the solar flux at the equator $\ang{0.23}$ before sunset and near the poles ($\lambda = \ang{80}$), $\ang{1.40}$ before sunset.
While the reflected flux is higher than the emitted longwave radiation, its influence is still much lower than flux directly coming from the sun.  Incorporating it into the model adds complexity without significantly improving the accuracy of the temperature estimates, hence it will be ignored in this study.

\paragraph{Eclipses}
While \citet{Ashkenazy2019} estimated that eclipses on Europa result in a general temperature decrease of only 0.3 K, we believe that some simplifications used there lead to an underestimation of the eclipse effects. Specifically, the model smooths the impact of eclipses over the entire diurnal period, resulting only in a slightly lower average solar flux (see Equation 30 of his article). In addition, this previous approach does not account for the fact that in some regions, eclipses occur during peak solar flux periods, causing brief but significant temperature drops. In this study, we investigate the effects of these eclipses through numerical simulations to more accurately consider their potential impact.

We have explored the effect of eclipses on Europa by integrating them into the input flux as a step function for a  duration of about 2.8 hours (see \citet{Ashkenazy2019}, parameter "$p$"). The timing of the eclipse is determined by the current longitude, for instance the middle of the eclipse is occurring at noon for longitude $\ang{0}$.  Here, we only account for the umbra effect and consider that the flux is zero during the eclipse, ignoring the penumbra, due to the considerable distance between Jupiter and the Sun, and Europa's proximity to Jupiter \citep{Ashkenazy2019}.
To estimate the maximum effect of eclipses we have chosen the potential warmest scenario on Europa, where we have the same thermal properties as previously but the albedo is now $A=0.4$ and we look at the equator where the solar flux is at his highest. 

Taking into account Europa's eclipses, that are short phenomena, requires a very precise timestep. Discontinuity in the input solar flux can lead to brief errors in the surface temperature, that will be dissipated better with a higher number of timesteps and improved boundary condition using Equation \ref{eq:linearizeT4}. While more advanced formulations of the upper boundary condition can be more robust to discontinuity \citep{Schorghofer2022}, for reasons of simplicity here, we choose a high number of $n_t=200$ timesteps per day to prevent appearance of errors.
Due to this higher resolution, we computed the eclipse histograms over a period of only 1000 years to keep computation time reasonable.

\begin{figure}[htpb]
	\centering
	\includegraphics[width=\textwidth]{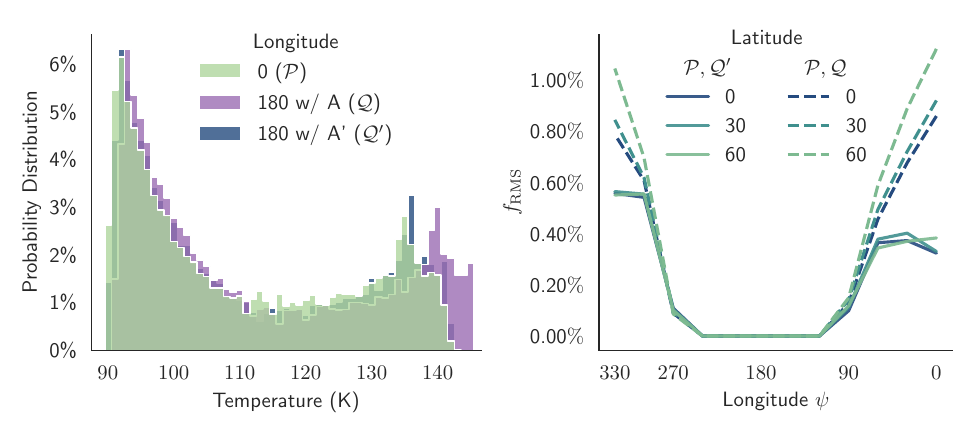}
	\caption{(\textit{Left}) Surface temperature distributions at Europa's equator for three different scenarios: $\mathcal{P}$ at longitude $\ang{0}$ (sub-Jovian, with eclipse) and $\mathcal{Q}$ at longitude $\ang{180}$ (anti-Jovian without eclipse) both with albedo $A=0.4$, and $\mathcal{Q'}$ at longitude $\ang{180}$ with equivalent albedo $A'=0.453$. 
    (\textit{Right}) RMS distance between eclipse temperature distributions $\mathcal{P}$ and non-eclipse distributions  with unchanged albedo $\mathcal{Q}$ or with equivalent albedo $\mathcal{Q'}$ as a function of longitude and latitude. Using an equivalent albedo allows to approximate, at first order, the surface temperature distribution of the sub-Jovian hemisphere without having to compute eclipses.
 }
	\label{fig:eclipse}
\end{figure}

In Figure (\ref{fig:eclipse} \textit{Left}, \textit{green histogram}), we have plotted the distribution of the surface temperatures  for a surface at latitude $\ang{0}$ and longitude $\ang{0}$, over 1000 years (excluding the initial convergence).
Naturally, with reduced solar flux, there is a noticeable temperature decrease, up to $\SI{5}{K}$, compared to the anti-Jovian hemisphere unaffected by eclipses (see Figure \ref{fig:eclipse} \textit{Left}, \textit{purple histogram}).
On the leading sub-jovian hemisphere, eclipses occur during the morning, stopping the surface heating and thereby reducing the maximum daily temperature. The general shape of the distribution (not represented here) is similar to the non-eclipse histogram (Figure \ref{fig:eclipse} \textit{Left}, \textit{purple histogram}), but shifted towards lower temperatures. 
On the trailing sub-jovian hemisphere, eclipses occur during the afternoon, after the peak daily solar flux. The surface has been heated enough to reach the anti-Jovian maximum temperature but only for a short duration. These histograms (not represented here) are characterized by a smaller second peak at mid temperatures, that is the temperature at which the surface drops during the eclipse. 

\paragraph{Equivalent Albedo}
To run simulations on the million year timescale within a reasonable computation time, we propose to introduce an equivalent albedo that approximates the effect of eclipses on surface temperature distributions. 
First, we consider a surface at longitude $\ang{180}$ with albedo $A$ and temperature at depth $T_{\infty} = T_{n_x-1}$, that is on the anti-Jovian hemisphere, not experiencing any eclipses.
Following \citet{Spencer1989}, the temperature at depth is proportional to the sub-solar temperature $T_{\mathrm{SS}}$:
\begin{equation}
    T_{\infty} = g(\Theta) T_{\mathrm{SS}}(A)
    \label{eq:TinftyTSS}
\end{equation}
where $g$  is the function given by the "$T_{\mathrm{DEEP}}$" curve on Figure 2 of \citet{Spencer1989}  and $\Theta$ is the thermal parameter defined by:
\begin{equation}
\Theta = \dfrac{\Gamma \sqrt{2 \pi/p_{\mathrm{E}}}}{(\epsilon \sigma)^{1/4}} \left( \dfrac{d^2}{(1-A) G_{\mathrm{SC}} } \right)^{3/4}
\end{equation}
where for Europa's thermal inertia and albedo range, $\Theta > 1$, classifying it as a fast rotator.
In contrast, a surface on the sub-Jovian hemisphere is subject to eclipses, so it receives less solar energy, resulting in a colder temperature at depth $T_{\infty}' < T_{\infty}$. 
To address this, we propose approximating such surface by an equivalent surface that does not experience eclipses ($\psi=\ang{180}$) but has a modified albedo $A'>A$ such that the temperature at depth matches $T_{\infty}'$.
Using Equation \ref{eq:TinftyTSS} and developing the expression of the sub-solar temperature \citep{Spencer1989}, this leads to the expression of the temperature at depth:
\begin{equation}
T_{\infty}' = g(\Theta') \left(  (1-A') \dfrac{G_{\mathrm{SC}}}{ \epsilon \sigma d^2} \right)^{1/4}.
\label{eq:Tinfty_dev}
\end{equation}
Although $\Theta$ is a function of $A$, in the case $\Theta>1$, the variations of $g(\Theta(A))$ with $A$  are small enough to be considered constant at first order, so $g(\Theta') = g(\Theta)$. 
Using the ratio of Equation \ref{eq:Tinfty_dev} for the two presented cases, leads to an expression of the equivalent albedo:
\begin{equation}
     A' = 1- (1-A) \left( \dfrac{T_{\infty}'}{T_{\infty}} \right)^4 
     \label{eq:equi_albedo}
\end{equation}
which can be computed knowing the original albedo $A$, the reference ($\psi = \ang{180}$) temperature at depth $T_{\infty}$ and the temperature at depth $T_{\infty}'$ of the surface subjected to an eclipse 

Using our numerical model, a parameter exploration was run on the longitudes ranging from $\ang{0}$ to $\ang{360}$ with a step of $\ang{30}$, and latitudes $\ang{0}$, $\ang{30}$, $\ang{60}$ (due to symmetry), for an albedo of $A=0.4$.
This allows us to obtain the temperatures at depth $T_{\infty}'(\lambda, \psi)$, notably for each of the sub-Jovian longitudes. 
Using Equation \ref{eq:equi_albedo}, we obtain an equivalent albedo, $A'= 0.433$ for longitudes $\ang{60}$ and $\ang{300}$, $A' = 0.449$ for longitudes $\ang{30}$ and $\ang{330}$ and $A'=0.453$ at longitude $\ang{0}$. No noticeable changes is observed with latitude.
The borders of the sub-Jovian hemisphere (longitudes $\ang{90}$ and $\ang{270}$) only experience half the duration of an eclipse respectively at the very beginning and end of the day, leading to an almost identical temperature distribution than on the anti-Jovian hemisphere.

As shown by Figure \ref{fig:eclipse}, at first order, the temperature distribution $\mathcal{P}$  in the sub-Jovian hemisphere (\textit{green histogram}) can be approached using an equivalent albedo $A'$ at longitude $\ang{180}$ (\textit{blue histogram}). 
The temperature distribution $\mathcal{Q}'$  using the equivalent albedo is considerably closer to reality than the distribution $\mathcal{Q}$ where albedo remains unchanged (\textit{purple histogram}) . To quantify the approximation error, we introduce the root mean square (RMS) distance between the two distributions:
\begin{equation}
    f_{\mathrm{RMS}}(\mathcal{P}, \mathcal{Q}) = \sqrt{  \dfrac{1}{n_{\mathrm{B}}} \sum_{b \in \mathrm{B}} \left(  \mathcal{P}(T_{b}) - \mathcal{Q}(T_{b}) \right)^2  }
\end{equation}
where $B$ is the set of bins, $n_{\mathrm{B}}$ is the number of bins and $T_b$ the temperature of bin $b$. 

First, the RMS distance is computed between eclipse-inclusive simulations distributions $\mathcal{P}$ and non-eclipse simulations distributions $\mathcal{Q}$ with the same albedo $A$ (\ref{fig:eclipse} \textit{Right}, \textit{dotted}). 
The mean RMS for the sub-Jovian leading hemisphere is $0.57\%$, for the sub-Jovian trailing hemisphere is $0.70\%$, and at longitude $\ang{0}$ is $0.86\%$.
Then, the RMS distance is computed  between eclipse-inclusive  simulations distributions $\mathcal{P}$ and the non-eclipse distributions using equivalent albedos $\mathcal{Q'}$ (\ref{fig:eclipse} \textit{Right}, \textit{lines}). 
In this case, the mean RMS for the sub-Jovian leading hemisphere is $0.37\%$, for the sub-Jovian trailing hemisphere is $0.51\%$, and at longitude $\ang{0}$ is $0.32\%$. Errors are higher on the sub-Jovian trailing hemisphere due to the second peak at mid-latitudes. 

Overall, the results show that the RMS distance is consistently smaller for the equivalence approximation, particularly effective at approximating the eclipse scenario at longitude $\ang{0}$ where eclipses significantly reduce the solar flux. In this case, the errors are divided by a factor of more than 3. With this approximation, the errors are limited to 0.5\% in the RMS distribution with 1K-wide bins, meaning that on average only 0.5\% of the time spent on a particular temperature set (bounded by $\pm \SI{0.5}{K}$) is incorrect.
Such simplification allows to extrapolate the results of our simulations to the sub-Jovian hemisphere, with limited errors and without requiring the high resource costs associated with eclipse modeling.
While this equivalence is not perfect, it allows to run longitude-independent simulations, which are highly useful for multiphysics coupling, as discussed in the joint publication \citep{Mergny2024i}.

\section{Conclusion and Perspective}

We have developed an efficient open-source fully implicit algorithm called MultIHeaTS, which uses finite differences to solve the heat equation on 1D heterogeneous media with an irregular grid. While our primary focus is planetary science, our algorithm is adaptable and can accommodate different types of boundary conditions and surfaces.

For homogeneous cases, the algorithm was validated using a known analytical solution.
This validation which used a discontinuous initial condition showed the robustness of MultIHeaTS to stiff conditions, in contrary to the Crank-Nicolson method.
For heterogeneous cases, MultIHeaTS was validated by comparison to a well established explicit algorithm in planetary science \citep{Spencer1989}.
Our fully implicit scheme remained accurate and stable, producing results that closely matched Spencer's explicit model.
Thanks to its advantageous stability, MultIHeaTS can compute the same temperature for a given time up to 100 times faster than Spencer’s explicit method. This is particularly advantageous for simulating processes that occur on large timescales.
Future research should conduct a detailed stability analysis to accurately quantify the limitations of the linearization of temperature-dependent properties and the Stefan-Boltzmann equation.

The capability of MultIHeaTS to simulate complex media were illustrated by computation of a bilayer surface. Based on our findings, we recommend an observation strategy: measuring the surface temperature at the morning sunrise, at noon and at midnight to characterize between bilayer and homogeneous surface profiles.
A second application took advantage of the fast computation capability of MultIHeaTS by estimating the temperature profile of an anti-Jovian surface on Europa during one million year, which includes the diurnal variations and the variation of the orbital element of the Galilean moon. This enabled us to draw for the first time the temperature distribution with depth on such a large timescale.
Finally, the effect of Jupiter's eclipses on the sub-Jovian hemisphere was explored, enabling us to propose an equivalence using a modified albedo with limited errors.
MultIHeaTS has a wide range of potential applications and is notably at the core of our multi-physics simulations to study ice microstructure presented in the joint article \citep{Mergny2024i}.
Thanks to its finite-difference formulation, the solver can be easily coupled with other processes, such as sintering, or phase change.

%The MultIHeaTS algorithm has a wide range of potential applications, and its open-source and flexible nature coupled with its computational efficiency make it adaptable to various domains. 

%Solvers like MultIHeaTS are expected to be useful in planetary science for retrieving thermal properties of multi-layered planetary surfaces using brightness temperature data.

%Our comparison of thermal signatures between homogeneous and bilayer profiles highlights the importance of using multilayer solvers to accurately model and analyze the thermal characteristics of planetary surfaces. Based on our findings, we recommend conducting observations during upcoming missions to Europa at three distinct local times: noon, midnight, and the end of the night. This approach will enable differentiation between bilayer and homogeneous profiles, enhancing our ability to accurately characterize the surface composition and thermal properties of Europa.

%Due to its finite-difference formulation, the solver can be easily coupled with other physical processes, as in the joint article where it is used to study the evolution of planetary icy surfaces microstructure.

\appendix

\section{Dirichlet Boundary Conditions}
\label{sec:app_dirichlet}
The temperature is fixed at the boundaries which gives us at the top boundary $x=0$:
\begin{equation}
    \forall t, T(0, t) = T_0(t) \implies
    \begin{cases}
        &b_0^i = 1 \\
        &c_0^i = 0 \\
        &s_0^i = T_0^i, \\
    \end{cases}  
\end{equation}
at at the bottom boundary $x=L$,
\begin{equation}
    \forall t, T(L, t) = T_{nx-1}(t) \implies
    \begin{cases}
        &a_{n_x-1}^i = 0 \\
        &b_{n_x-1}^i = 1 \\
        &s_{n_x-1}^i = T_{n_x-1}^i. \\
    \end{cases}  
\end{equation}

\section{Stability and Accuracy of the Finite Difference Schemes in Standard Conditions}
\label{sec:stability}

Before analyzing the behavior of finite difference schemes in heterogeneous media and non-linear boundary conditions, it is instructive to perform a stability analysis in simpler homogeneous media and classical linear conditions.
We follow the Fourier transform approach as described in \citep{Pearson1965, Giles1995, Thomas1995, Oesterby2003, Langtangen2017}: by injecting wave components of different frequencies $\nu \in [-\pi, \pi]$ into the heat Equation \ref{eq:heat-equation}, we can look at the response and behavior of the different schemes.

The propagation of the finite difference solution from one time step to the next is governed by the growth factor which for the explicit method  \citep{Press1992} is
\begin{equation}
    g(\nu) = 1 - 4 F \sin^2\dfrac{\nu}{2}
\end{equation}
where values of $\nu$ close to 0 indicate low frequency components and values close to $\pi$ indicate high oscillatory wave components and  $F$ is the Fourier number defined as
\begin{equation}
    F = \dfrac{\alpha \Delta t }{ \Delta x^2}
\end{equation}
The growth of numerical wave components occurs when the absolute value of $g(\nu)$ exceeds 1,  which for explicit scheme happens if $F > \frac{1}{2}$. This leads to the magnification of the corresponding wave component, indicating instability.
Consequently, the explicit scheme is numerically unstable, and its domain of stability is given by
\begin{equation}
   F < 1/2 
   \label{eq:stability}
\end{equation}
The same analysis \citep{Press1992} for the Backward Euler scheme leads to a growth factor of
\begin{equation}
    g(\nu) = \left( 1 + 4 F \sin^2 \dfrac{\nu}{2}   \right)^{-1}
\end{equation}
In this case, the growth factor is constrained by $\left| g(\nu) \right| \le 1$ for all $F$ and $\nu$, meaning that the fully implicit scheme is unconditionally stable.

Finally the growth factor of the Crank-Nicolson method \citep{Press1992} is given by
\begin{equation}
    g(\nu) = \dfrac{1- 2 F \sin^2 \frac{\nu}{2}}{1+2F \sin^2 \frac{\nu}{2}}.
\end{equation}
It is clear that the Crank-Nicolson method is also unconditionally stable as $\left| g(\nu) \right| \le 1$ for all values of $F$ and $\nu$.
However we note that for high frequency components $\nu \to  \pm \pi$ then $g(\nu) \to -1$, especially for large $F$, meaning that these components are propagated as weakly damped oscillations.

Notably, the growth factor of the fully implicit scheme is relatively small for high Fourier Numbers, which means that the components that exhibit the most problematic behavior in the Crank-Nicolson method are the same components that are most effectively damped by the fully implicit method \citep{Oesterby2003}.  Thus, the fully implicit scheme has an advantage of stability when handling stiff initial conditions, which is illustrated in the results section.

\section*{Statements and Declarations}
\subsection*{Funding and Competing Interests}

We acknowledge support from the ``Institut National des Sciences de l'Univers'' (INSU), the ``Centre National de la Recherche Scientifique'' (CNRS) and ``Centre National d'Etudes Spatiales'' (CNES) through the ``Programme National de Plan{\'e}tologie''. 
We thank the two anonymous reviewers for their careful reading of our manuscript and their insightful comments and suggestions. We also extend our gratitude to Dr. Alice Le Gall and Salman Raza for testing the solver and providing valuable guidance on accommodating nonlinear boundary conditions.

The authors have no competing interests to declare that are relevant to the content of this article.

\section*{Data Availability}
The authors declare that the data supporting the findings of this study are available within the paper and online at the IPSL Data Catalog:\dataset[10.14768/9763d466-db02-4f29-8ad5-16e6e0187bd4]{\doi{10.14768/9763d466-db02-4f29-8ad5-16e6e0187bd4}}

%\newpage
%\section*{Figures}

\bibliography{library}
\bibliographystyle{aasjournal}

\end{document}

%\articletype{ARTICLE TEMPLATE}% Specify the article type or omit as appropriate

\title{Supplementary Materials - Bilayer Surface Characterization Strategy from Space using MultIHeaTS: A Fast and Stable Thermal Solver }

\author*[1]{\fnm{C.} \sur{Mergny}}\email{cyril.mergny@universite-paris-saclay.fr}
\author[1,2]{\fnm{F.} \sur{Schmidt}}\email{frederic.schmidt@universite-paris-saclay.fr}

\affil*[1]{\orgdiv{GEOPS}, \orgname{Paris-Saclay University}, \city{Orsay}, \country{France}}
\affil[2]{ \orgname{Institut Universitaire de France}, \city{paris}, \country{France}}

\maketitle

\section{Dirichlet Boundary Conditions}
The temperature is fixed at the boundaries which gives us at the top boundary $x=0$:
\begin{equation}
    \forall t, T(0, t) = T_0(t) \implies
    \begin{cases}
        &b_0^i = 1 \\
        &c_0^i = 0 \\
        &s_0^i = T_0^i, \\
    \end{cases}  
\end{equation}
at at the bottom boundary $x=b$,
\begin{equation}
    \forall t, T(b, t) = T_{nx-1}(t) \implies
    \begin{cases}
        &a_{nx-1}^i = 0 \\
        &b_{nx-1}^i = 1 \\
        &s_{nx-1}^i = T_{nx-1}^i. \\
    \end{cases}  
\end{equation}

\section{Analytic Solution of a Step Fonction}
\label{sup:ana}
\subsubsection{Fourier series}
Any periodical function $f$ in $\mathbb{R}$ of period $P$ can be written as a Fourier series:
\begin{equation}
    f(x) =  \frac{a_0}{2} + \sum_{n=1}^{+\infty} a_{nx-1} \cos\left(\frac{2\pi n x}{P}\right) + b_n \sin\left(\frac{2\pi n x}{P}\right),
    \label{eq:fourier_series}
\end{equation}
with the coefficients $a_n$ and $b_n$ given by the expressions:
\begin{equation}
    a_n = \frac{2}{P} \int_P f(x) \cos\left(\frac{2\pi n x}{P}\right) dx,
    \label{eq:a_n}
\end{equation}
\begin{equation}
    b_n = \frac{2}{P} \int_P f(x) \sin\left(\frac{2\pi n x}{P}\right) dx.
    \label{eq:b_n}
\end{equation}
The expressions are obtained by integrating equation (\ref{eq:fourier_series}) as demonstrated here:
\begin{align*}
    \int_P f(x) \cos\left(\frac{2\pi m x}{P}\right) dx & =  \int_P  \cos\left(\frac{2\pi m x}{P}\right)\left[ \frac{a_0}{2} + \sum_{n=1}^{+\infty} a_n \cos\left(\frac{2\pi n x}{P}\right) + b_n \sin\left(\frac{2\pi n x}{P}\right) \right] dx \\
    &=  \sum_{n=1}^{+\infty} \int_P  a_n \cos\left(\frac{2\pi n x}{P}\right)  \cos\left(\frac{2\pi m x}{P}\right) dx \\
    &= a_m \frac{P}{2}
\end{align*}

\subsubsection{Fourier series of a step function}
Every function on a interval of length $P$ can be extended to a periodic function in $\mathbb{R}$ of period $P$.
First we will extend function $H$ on the interval $[-\frac{L}{2}, \frac{3L}{2}]$ such that the derivative at $x=0$ and $x=L$ are equal to zero (Boundary flux condition). 
The step function is defined on the interval $[-\frac{L}{2}, \frac{3L}{2}]$.

We use equations (\ref{eq:a_n}) and (\ref{eq:b_n}) with $P=2L$ the interval length to find the coefficient $a_n$ and $b_n$:
\begin{align*}
    a_n &= \frac{1}{L} \int_{-\frac{L}{2}}^{\frac{3L}{2}} H(x) \cos\left(\frac{\pi n x}{L}\right) dx \\
    & = \frac{1}{L} \int_{\frac{L}{2}}^{\frac{3L}{2}} \cos\left(\frac{\pi n x}{L}\right) dx \\
    & = \frac{-2}{\pi n} \sin\left( \frac{\pi n}{2} \right)
\end{align*}
and a similar integration for $b_n$ leads to $b_n=0$.
The step function can then be written as a Fourier series using equation (\ref{eq:fourier_series}):
\begin{equation}
    H(x) = \frac{1}{2} - \sum_{n=1}^{+\infty} \frac{4}{\pi n} \sin\left(\frac{\pi n }{2} \right) \cos\left(\frac{\pi n x}{L}\right).
\end{equation}

\subsubsection{Solving the heat equation by separation of variables}

The solution of the 1D heat equation  with a step function initial condition, can be obtained by separation of variables. Let's assume that the solution can be written as
\begin{equation}
    T(x, t) = v(x)w(t)
\end{equation}
with $v$ and $w$ functions of $x$ and respectively $t$. This expression injected int the heat equation results to:
\begin{equation}
   v(x) \dfrac{\partial w(t)}{\partial t} = \alpha  \dfrac{\partial^2 v(x)}{\partial x^2} w(t)
\end{equation}
This first order differential equation as for $w$ the general solution:
\begin{equation}
    w(t) = w(0) e^{\alpha \dfrac{v^{''}(x)}{v(x)} t}
\end{equation}
where $v^{''}$ is the second derivative of $v$ with respect to $x$. If we take $v(x) = T_n(x, 0) = f_n(x)$ the initial function decomposed in 
a Fourier series, then we have:
\begin{equation}
        v^{''}(x) = -\left(\frac{2 \pi n}{P} \right)^2 f_n(x)
\end{equation} 
and 
\begin{align*}
    & T(x, 0) = v(x)w(0) = f(x) \\
    & v(x) = f(x) \implies w(0) = 1
\end{align*}
which give us the exact express of the solution of $w$ with:
\begin{equation}
    w(t) = e^{-\alpha \left(\dfrac{2 \pi n}{P} \right)^2 t}
\end{equation}
Hence the solution is given by:

\begin{equation}
    T(x, t) = \frac{a_0}{2} + \sum_{n=1}^{+\infty} \left(a_n \cos\left(\frac{2\pi n x}{P}\right) + b_n \sin\left(\frac{2\pi n x}{P}\right) \right)  e^{-\alpha \left(\dfrac{2 \pi n}{P} \right)^2 t}
\end{equation}

%\subsection{Parameters}

%\bibliography{library}
%\bibliographystyle{tfq.bst}